\documentclass[mathleft
]{an}
\usepackage{graphicx}
\usepackage{float}
\usepackage{times}
\begin{document}

\Pagespan{789}{}
\Yearpublication{2006}%
\Yearsubmission{2005}%
\Month{11}%
\Volume{999}%
\Issue{88}%

\title{Theoretical study of $\gamma$ Doradus pulsations in pre-main sequence stars}

  \author{M.-P. Bouabid
          \inst{1,2}
          \and
          J. Montalb\'an\inst{2}
	  \and
	  A. Miglio\inst{2}
	  \and
	  M.-A. Dupret\inst{2}
	  \and
	  A. Grigahc\`ene\inst{3}
	  \and
	  A. Noels\inst{2}
	}

   \institute{UMR 6525 H. Fizeau, UNS, CNRS, OCA, Campus Valrose, F-06108 Nice Cedex 2, France\\
              \email{bouabid@oca.eu}
	 \and
              Institut d'Astrophysique et de G\'eophysique de l'Universit\'e de Li\`ege, All\'ee du 6 Ao\^ut, 17 B-4000 Li\`ege, Belgium\\
	 \and
	     Centro de Astrofisica da Universidade do Porto, Rua das Estrelas, 4150-762 Porto, Portugal\\
             }

\received{30 May 2005}
\accepted{11 Nov 2005}
\publonline{later}

\keywords{stars: variables: gamma Doradus stars - stars: oscillations - stars: pre-main sequence}

\abstract{%
The question of the existence of pre-main sequence (PMS) $\gamma$~Doradus ($\gamma$~Dor) has been raised by the observations of young clusters such as NGC~884 hosting $\gamma$~Dor members.
We have explored the properties of $\gamma$~Dor type pulsations in a grid of PMS models covering the mass range $1.2 M_\odot < M_* < 2.5 M_\odot$ and
we derive the theoretical instability strip (IS) for the PMS $\gamma$~Dor pulsators. 
We explore the possibility of distinguishing between PMS and MS $\gamma$~Dor by the behaviour of the period spacing of their high order $gravity$-modes ($g$-modes).
}

\maketitle

\section{Introduction}

   The variability of  $\gamma$~Dor stars was identified as due to pulsations by Balona, Krisciunas $\&$ Cousins (1994), and the features of the stars belonging to 
   this new class of pulsators were defined by Kaye et al. (1999).   
   They are late A and F-type stars covering a part of the Hertzsprung-Russel diagram (HRD) between $7200-7700~K$ on the zero-age main sequence (ZAMS) and $6900-7500~K$ above it (Handler 1999), between 
   the solar-like stars and the $\delta$~Scuti ($\delta$~Sct) IS.
   They are located between stars with a deep convective envelope (CE) and stars with a radiative envelope, in the region of the HRD where 
   the depth of the CE changes rapidly with the effective temperature of the star.
   $\gamma$~Dor pulse with high order $g$-modes in a range
   of periods between 0.3-3 days. The excitation mechanism proposed by Guzik et al. (2000)
   using the frozen-convection assumption is a modulation of the radiative flux at the base of the CE.
   This mechanism was revisited by 
   Dupret et al. (2005) using a time-dependant convection (TDC) treatment.  

   From an observational point of view, the limits of the $\gamma$~Doradus IS have been lastly established by 
   Handler $\&$ Shobbrook (2002) (HS02 hereafter) and in the rest of the paper these limits will be adopted to define the $\gamma$ Dor IS.
   Since the depth of the CE plays a major role in the driving mechanism of $\gamma$~Dor pulsations, the theoretical predictions of stability
   are very sensitive to the parameter $\alpha_{MLT}$ defining the travel length of convective elements in the classical mixing-lenght treatment (MLT) of convection (B\"ohm-Vitense 1958). 

\begin{figure}[!h]
\centering
\includegraphics[width=70mm,height=60mm]{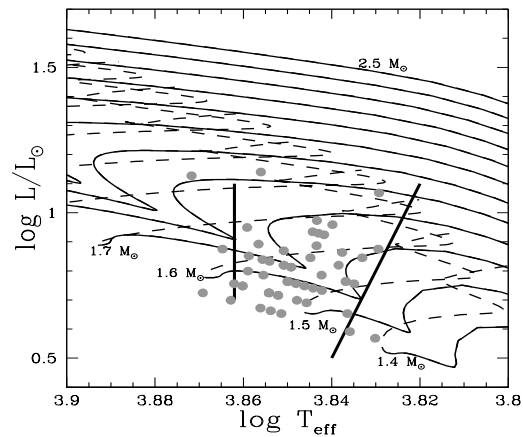}
\caption{PMS and MS (resp. full $\&$ dashed thin lines) evolutionary tracks. Thick straight lines correspond to the limits of the observational $\gamma$~Dor IS (HS02). Grey points represent the $bonafide$ $\gamma$~Dor stars from Henry, Fekel $\&$ Henry (2005).}
\label{HRdiag}
\end{figure}

   Number of observational efforts have been devoted to the search of PMS $\gamma$~Dor pulsators. 
   Saesen et al. (2010) found 6 multiperiodic A and F-type stars with a mean frequency between $0.2$~d$^{-1}$ and $3$~d$^{-1}$ during their multisite observation campain on the young open cluster
   NGC~884 (age $\sim$ 12.8 Myr - Slesnick et al. 2002).
   Zwintz et al. (2009) searched for PMS pulsators in a young open cluster (NGC~2264, 
   age $\sim$ 3-10 Myr - Sung et al. 2004; Sagar et al. 1986) but did not find any $\gamma$~Dor which were confirmed cluster members.

   The above mentioned theoretical works on $\gamma$~Dor stars systematically studied MS models stability. 
   However, as shown in Fig.~\ref{HRdiag}, MS and PMS\footnote{we consider as PMS models those before the onset of the stationary central H-burning} 
   evolutionary tracks cross the observational $\gamma$~Dor IS.
   The presence of stars at different evolutionary phases in this region of the HRD raises some questions: since the internal structure of PMS stars is different,
   can we expect $\gamma$~Dor pulsations in such stars? Could these pulsations be used to distinguish between PMS and MS $\gamma$~Dor?

   To answer these questions, we performed an adiabatic and a non-adiabatic asteroseismic analysis on a grid of PMS and MS models between $1.2$ and $2.5 M_{\odot}$ computed with the stellar evolution code 	CLES (Scuflaire et al. 2008a).
   The adiabatic and non-adiabatic computations have been done respectively with the LOSC (Scuflaire et al. 2008b) and MAD (Dupret 2001) codes.

\begin{figure}[!hb]
\centering
\includegraphics[width=60mm,height=40mm]{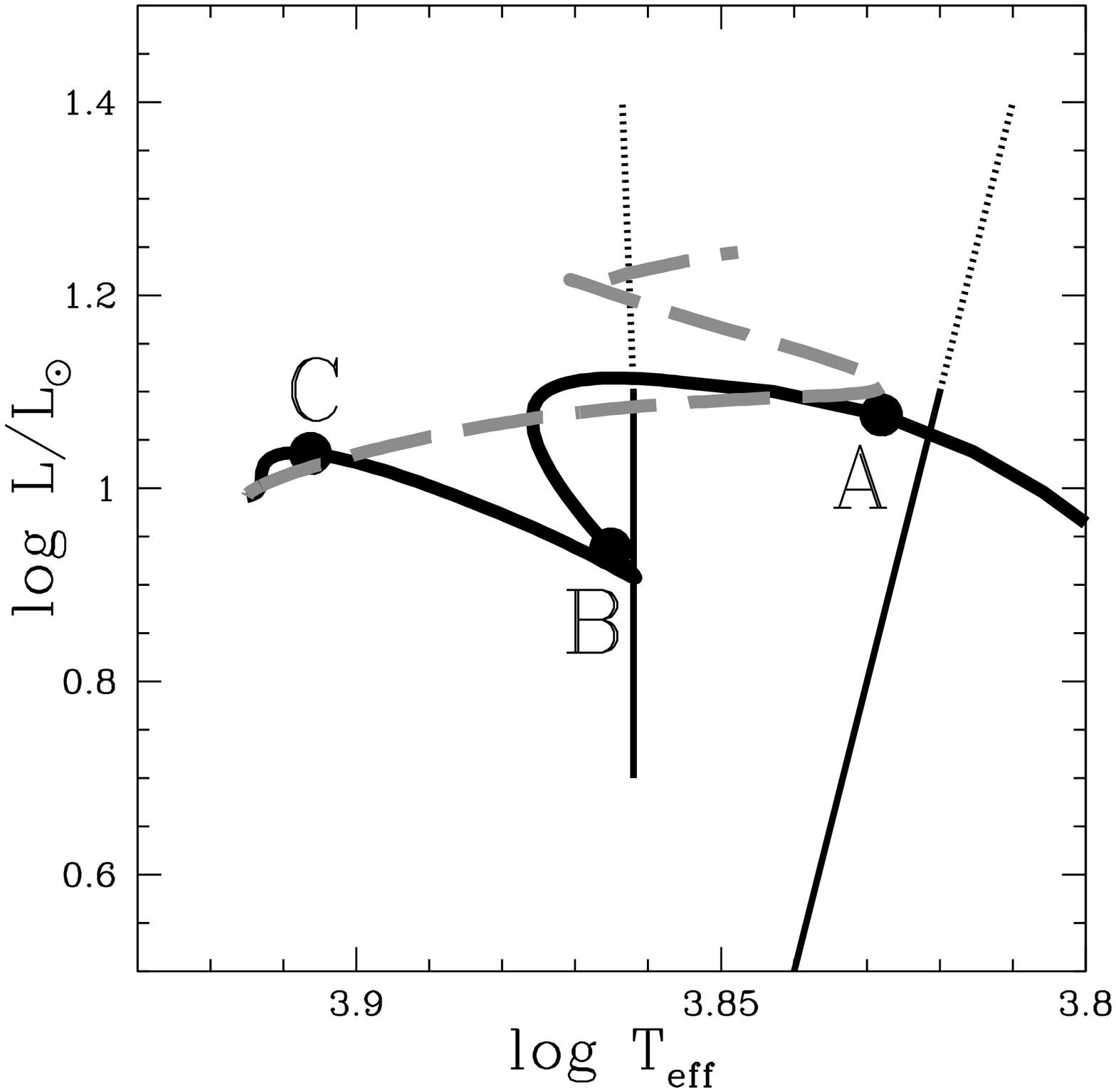}
\includegraphics[width=60mm,height=40mm]{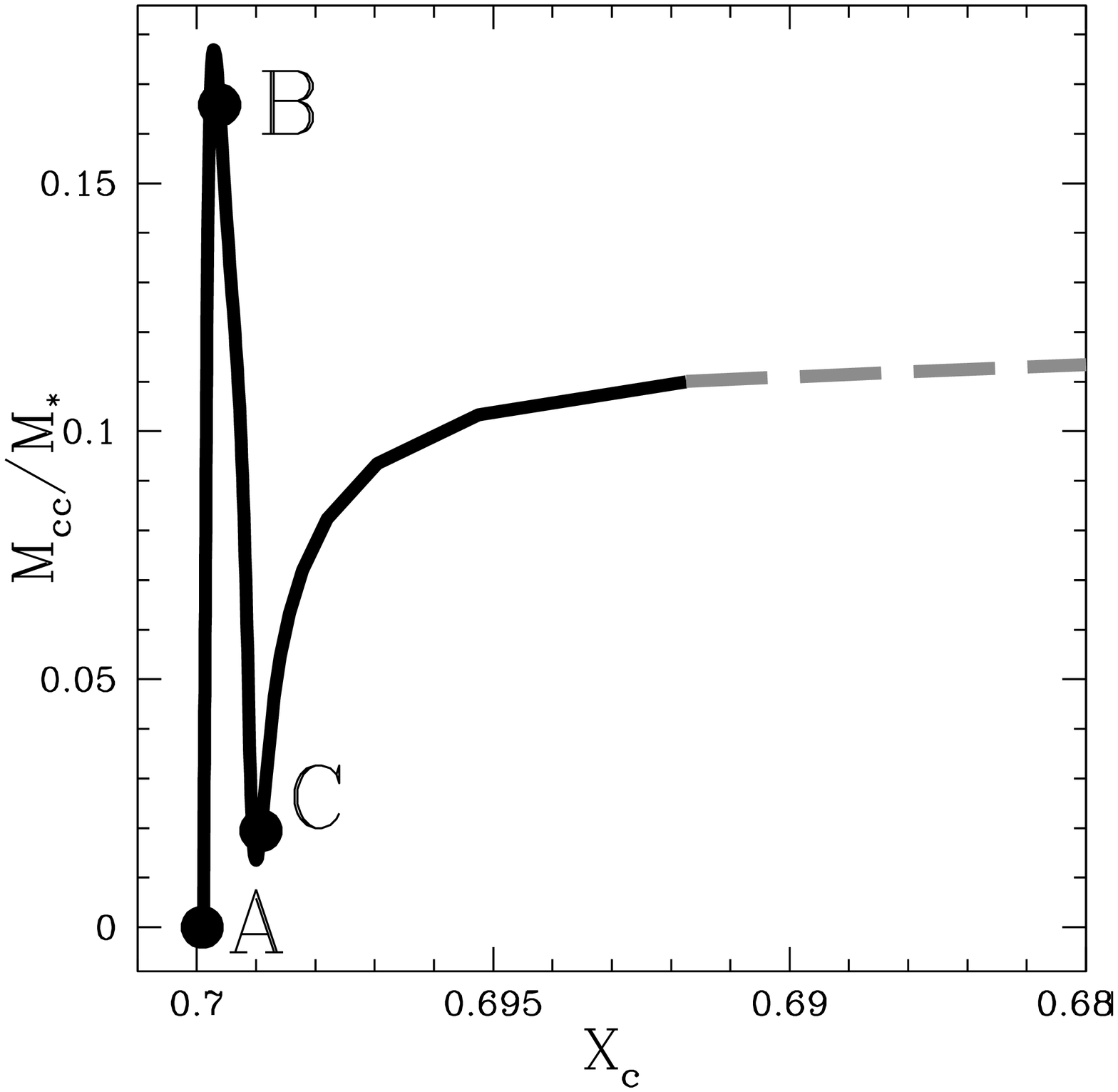}
\includegraphics[width=60mm,height=40mm]{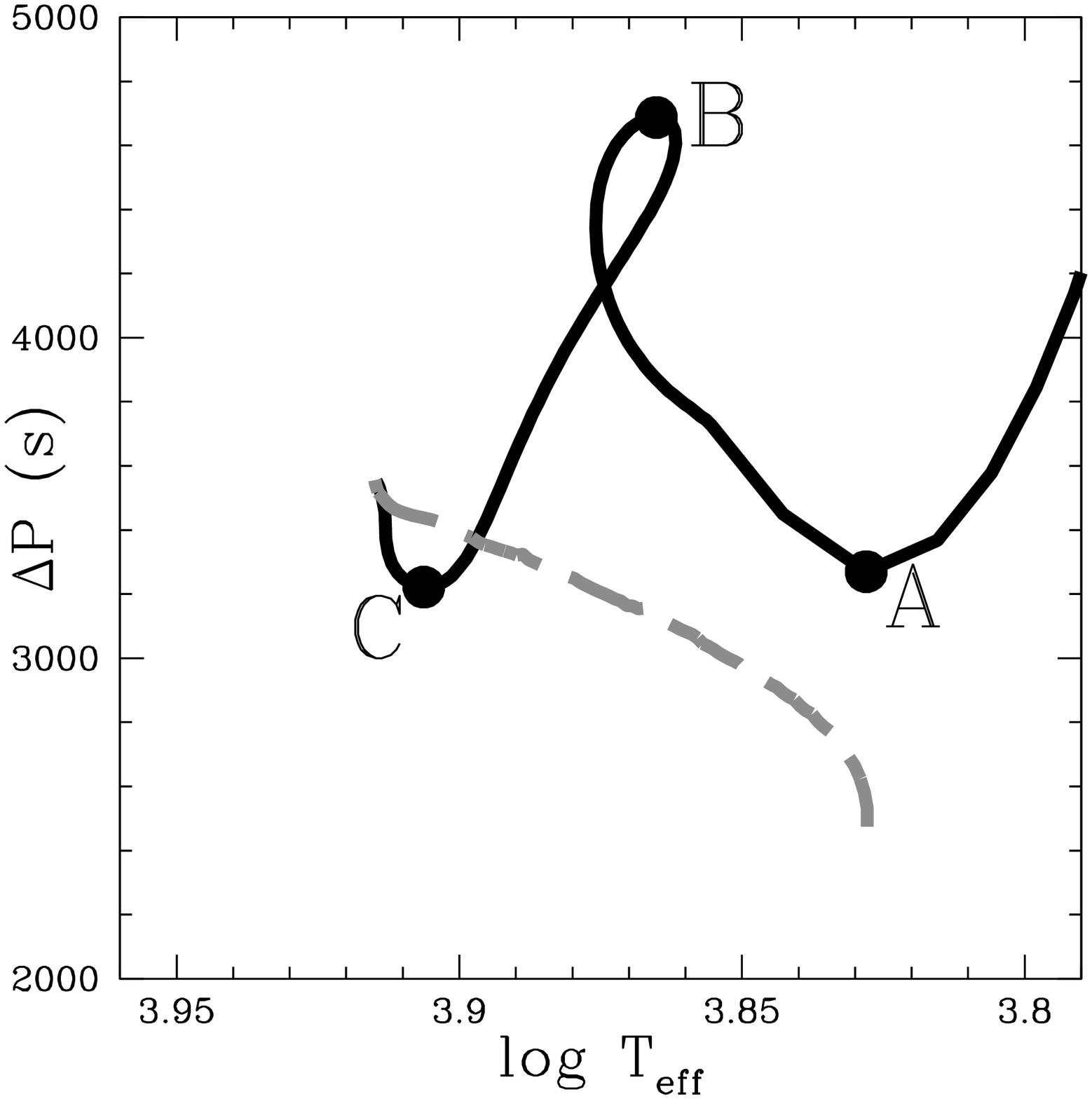}
\caption{
Top panel: Evolution of a $1.8 M_\odot$ star in the HRD (PMS phase in full black and MS phase in dashed grey).
Middle panel: Variation of its CC mass from the PMS phase to the early MS phase ($X_c = 0.68$).
Bottom panel: Evolution of the $\ell = 1$ modes period spacing as a function of the effective temperature of the star from the PMS to the Terminal Age Main Sequence (TAMS).
}
\label{periodspaceall}
\end{figure}


\section{Internal structure and adiabatic study}

\begin{figure}[!h]
\centering
\includegraphics[width=70mm,height=60mm]{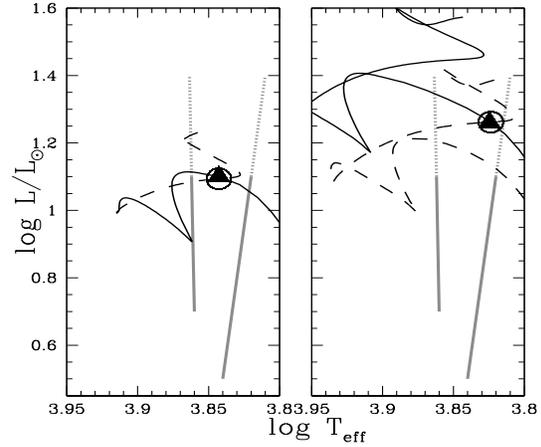}
\caption{Evolutionary tracks of different models crossing the observational $\gamma$~Dor IS (thick grey lines - HS02). Left panel: $1.8 M_\odot$ evolutionary track whose PMS (full line) and MS (dashed line) phases intersect inside the IS (circle: PMS model - triangle: MS model). Right panel: Evolutionary tracks for $1.9$ and $2.1 M_\odot$ showing the same HRD location for a $2.1 M_\odot$ PMS model (circle) and a model of $1.9 M_\odot$ at the end of its MS phase (triangle).}
\label{HRcompare}
\end{figure}

A star approaching the MS from the Hayashi track has already a radiative core that continues to contract. The increase of density due to this contraction leads to  the increase of the central temperature ($T_c$). This phenomenon continues until $T_c$ is high enough ($\sim 1.7~10^7 K$) to start the  nuclear reactions of the CN subcycle. Because of the high dependence on temperature of the $^{12}$C$(p,\gamma)^{13}$N$(\beta^+\nu)^{13}$C$(p,\gamma)^{14}$N nuclear reaction rate ($\propto T^{19}$) a convective core (CC) appears. The fraction of mass of this CC changes as the star evolves toward the ZAMS (Fig.~\ref{periodspaceall} - middle panel) and for typical $\gamma$~Dor stellar masses, the CC remains during the MS.
The onset of the CN subcycle appears in the PMS evolutionary track as a kind of loop with a minimum of luminosity after a first maximum (Fig.~\ref{periodspaceall} - top panel). 

As shown in Fig.~\ref{HRdiag}, the phase at which the star crosses the IS during the PMS changes with the stellar mass. While low mass models have already developed a CC, more massive models are still contracting with a fully radiative core.

The properties of the $g$-modes spectrum is determined by the matter stratification in the star, which is described by the Brunt-V\"ais\"al\"a frequency $N$:
\begin{equation}
N^2=g\left(\frac{1}{\Gamma_1}\frac{\textrm{d} \ln P}{\textrm{d}r}-\frac{\textrm{d} \ln \rho}{\textrm{d}r}\right)
\end{equation}
with $g$ the local gravity, $\rho$ the local density, $P$ the local pressure, $r$ the local radius and $\Gamma_1$ the first adiabatic exponent.

From the first order asymptotic theory (Tassoul 1980), the period of a $g$-mode with a radial order $k$ and a degree $\ell$ in a star with a CC and a CE is given by:
\begin{equation}
	P_k = \frac{\pi^2}{\sqrt{\ell(\ell+1)} \int_{r_1}^{r_2}{\frac{| N |}{r} \textrm{d}r}} \left( 2k + 1 \right)
\end{equation}
with $r_1$ and $r_2$ the limits of the $g$-mode cavity defined by $\sigma_g^2 < N^2, S_\ell^2$, where $\sigma_g$ is the $g$-mode frequency and $S_\ell$ the Lamb frequency for modes of degree $\ell$ (Fig.~\ref{BVfreq}). 
The dependence of $g$-mode periods on the behaviour of $N$ in the central region of the star allows us to expect a clear difference between the seismic properties of PMS and MS $\gamma$~Dor.

\subsection{Comparison of the models internal structures}

\begin{figure}[!h]
\centering
\includegraphics[width=80mm,height=60mm]{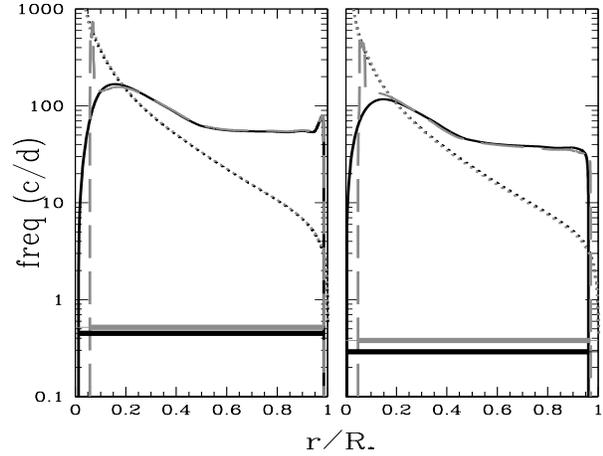}
\caption{Propagation diagram. PMS (black) and MS (grey) $N$ and "$\ell = 1$"-Lamb (dotted lines) frequencies for models with the same mass (left) and for models with different masses (right).
The straight thick lines represent the propagation of $g$-modes having the same radial order in the PMS (black) and MS (grey) models.}
\label{BVfreq}
\end{figure}

The PMS evolutionary track of a $1.8 M_\odot$ star crosses at different points the corresponding MS track. In particular, both tracks cross in the middle of the $\gamma$~Dor IS (Fig.~\ref{HRcompare} - left panel).  
By comparing these two models at the same location in the HRD we eliminate undesirable effects on the stellar structure coming from different effective temperatures or luminosities.
The internal structures of these PMS and MS models are shown in Fig.~\ref{BVfreq} where we plotted their $N$ and $S_{\ell = 1}$ frequencies as a function of the relative radius.

Because of the same radius and mass, PMS and MS models show similar behaviour of $N$ in the outer layers and the bases of the CE are located at the same depth (Fig.~\ref{BVfreq} - left panel).
Both $N$ profiles present also a bump in the inner layers, due to the density distribution. However, the central layers of the two models are very different. The PMS model has only a small CC ($N^2 < 0$) due to the onset of the CN subcycle while the CC of the MS model is larger. The main difference between the two $N$ profiles lies in the sharp feature located at the limit of the MS convective core. This peak is due to the presence of an important mean molecular weight gradient ($\nabla_\mu$) at the limit of a receding CC.

As mentioned above, more massive PMS models cross the $\gamma$~Dor IS during an earlier phase than lower mass ones. We considered models with CC overshooting in order to have MS models with high enough luminosity and low enough effective temperature crossing a more massive PMS track (Fig.~\ref{HRcompare} - right panel). A $2.1 M_\odot$ quasi chemically homogeneous PMS model that still has a radiative core has the same HRD location as a $1.9 M_\odot$ evolved MS model. The different masses but same radius of the two models lead to differences in their density profiles that determines the behaviour of $N$ in the envelope, and therefore the different depth of their CE (Fig.~\ref{BVfreq} - right panel).

\subsection{Adiabatic study - Period spacing}

From Eq. 2, the period spacing between two $g$-modes with consecutive radial orders and same degree can be written as:
   \begin{equation}
	\Delta P = P_{k+1} - P_{k} = \frac{2 \pi^2}{\sqrt{\ell(\ell+1)}\int_{r_1}^{r_2}{\frac{| N |}{r}dr}}
   \end{equation}

Fig.~\ref{periodspaceall} (bottom panel) represents the evolution of the period spacing from the PMS to the TAMS for 
a $1.8 M_\odot$ star. The period spacing presents a clear variation which is strongly dependant on the evolution of the CC.

For models with the same effective temperature, the maximum difference between PMS and MS period spacing is around 2000 seconds and corresponds to the maximum mass fraction of the CC (Fig.~\ref{periodspaceall} - point $B$).
The period spacing value could be, in principle, used as a first discriminant between PMS and MS $\gamma$~Dor. However the period spacing 
of PMS stars can be of the same order as the MS $\gamma$~Dor one (Fig.~\ref{periodspaceall} - bottom panel - points $A~\&~C$)
and we cannot always use this value to determine if a $\gamma$~Dor star is in its PMS or its MS evolutionary state.

\begin{figure}[!h]
\centering
\includegraphics[width=70mm,height=60mm]{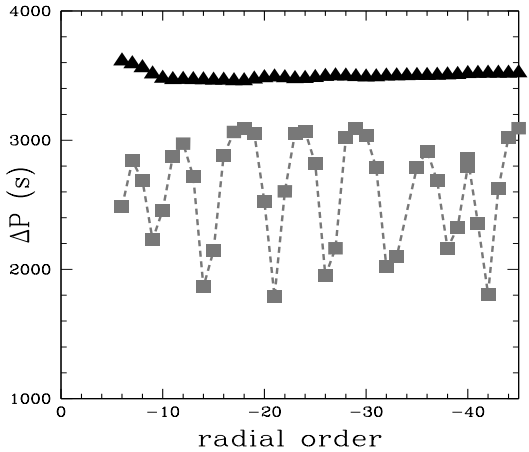}
\caption{Period spacing structure for $\ell=1$ modes as a function of the radial order of the modes for PMS (black triangles) and MS (grey squares) models with the same mass ($M = 1.8 M_\odot$) presented in subsection 2.1 and Fig.~\ref{HRcompare}.}
\label{periodspacestructure}
\end{figure}

Nevertheless, a sharp variation of $N$ such as the one due to the $\nabla_\mu$ at the border of the CC let a clear asteroseismic signature: the oscillation of the period spacing around its mean value (Miglio et al. 2008 
and references therein). While MS models can present that $\nabla_\mu$, PMS ones are almost chemically homogeneous, $i.e.$ their $N$ profile is quite smooth. Therefore the PMS period spacing does not clearly change with the radial order (Fig.~\ref{periodspacestructure}).


\begin{figure}[!h]
\centering
\includegraphics[width=70mm,height=60mm]{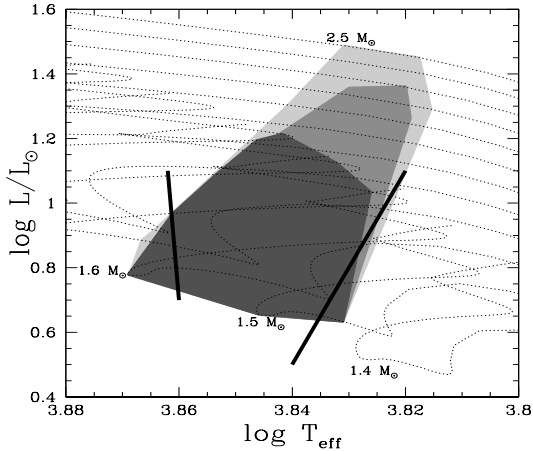}
\caption{$\gamma$~Dor theoretical IS for MS models with overshooting ($\alpha_{ov} = 0.20$, middle grey) and without (dark grey) and for PMS models (light grey). Straight lines correspond to the observational $\gamma$~Dor IS (HS02) and the thin dotted lines are the evolutionary tracks for models between $1.4$ and $2.5 M_\odot$.
}
\label{ISpms}
\end{figure}

\section{Non-adiabatic study - Stability analysis}

Because the adiabatic study does not take into account the excitation and damping of modes, we performed a non-adiabatic analysis.
The theoretical IS were calculated for $\ell=1$ and $\ell=2$ modes with the non-adiabatic code MAD including TDC treatment (Grigahc\`ene et al. 2005) on a grid of $1.2 - 2.5 M_\odot$ stellar models computed with the following phyical inputs:
\begin{itemize}
\item OPAL2001 equation of state (Rogers $\&$ Nayfonov 2002) and OP opacity tables (Badnell et al. 2005) completed at low temperature ($T < 10^4 K$) by the tables provided by Ferguson et al. (2005). 
The atmosphere tables were derived from Kurucz atmosphere models starting at the photosphere. We used the abundances from Asplund, Grevesse $\&$ Sauval (2005) for an initial metal mass fraction $Z = 0.02$ and an initial central hydrogen fraction $X = 0.70$. As we used the same non-adiabatic tools as Dupret et al. (2004), we chose the MLT parameter value they needed to tune their theoretical MS $\gamma$~Dor IS to the observational one, $i.e.$ $\alpha_{MLT} = 2.00$. They showed that a smaller value of the MLT parameter would shift the whole theoretical IS to lower effective temperatures. Finally, we computed models with and without convective core overshooting ($\alpha_{ov} = 0.20; 0.00$).
\end{itemize}

Our first results are presented in Fig.~\ref{ISpms} showing the theoretical IS of high order $g$-modes for MS and PMS models.
The location of PMS $\gamma$~Dor theoretical IS matches up with the MS one.


\section{Conclusion}

We carried out an adiabatic and a non-adiabatic studies on a grid of PMS and MS models in the mass range $1.2 M_\odot < M_* < 2.5 M_\odot$ to differenciate the 
asteroseismic behaviour of PMS $\gamma$~Dor from that of MS ones.

We pointed out the theoretical existence of PMS high order $g$-mode pulsators in the region of the observational $\gamma$~Dor IS. The theoretical PMS IS has the same edges than the MS one and presents a good agreement with the observational $\gamma$~Dor IS for $\alpha_{MLT} = 2.00$. 
The only difference between MS and PMS IS lies in the fact that even with a CC overshooting no MS evolutionary track can reach the upper region of the HRD, where unevolved massive PMS $\gamma$~Dor exist.

The measurement of the period spacing allows us to make the distinction between PMS and MS models by two different ways: 
\begin{itemize}
\item At fixed stellar parameters, the difference between central internal structures may lead to  a significant difference between the values of the period spacing. We are aware that we should investigate
if we can still distinguish PMS stars once we consider stellar parameters uncertainties but it is not the purpose of the present paper.
\item the behaviour of the period spacing is also different between MS and PMS models. While during the MS the important $\nabla_\mu$ and the evolution of the CC leads to an oscillation of the period spacing, the lack of such a $\nabla_\mu$ during the PMS phase leads to a period spacing independant of the mode radial order.
\end{itemize}

In a forthcoming paper the difference between PMS and MS non-adiabatic $g$-modes frequency spectra will be presented.


\end{document}